\documentclass[graybox]{svmult}

\usepackage{type1cm}        
\usepackage{makeidx}         
\usepackage{graphicx}        
\usepackage{multicol}        
\usepackage[bottom]{footmisc}

\usepackage{newtxtext}       %
\usepackage[varvw]{newtxmath}       

\usepackage{physics}
\usepackage{bm}
\usepackage{bbold} 
\usepackage{hyperref}
\hypersetup{
colorlinks=true,
citecolor=blue,
linkcolor=blue,
filecolor=blue,      
urlcolor=blue,
}

\newcommand{\PTr}[2]{\text{Tr}_{#1}\left\{#2\right\}}

\graphicspath{{figures/}}

\makeindex             

\begin{document}

\title*{Many-body thermal states on a quantum computer: a variational approach.}
\author{Mirko Consiglio and Tony J. G. Apollaro}
\institute{Mirko Consiglio \at Department of Physics, University of Malta, Msida MSD 2080, Malta, \email{mirko.consiglio@um.edu.mt }
\and Tony J. G. Apollaro \at Department of Physics, University of Malta, Msida MSD 2080, Malta \email{tony.apollaro@um.edu.mt}}
%
%
\maketitle

\abstract{Many-body quantum states at thermal equilibrium are ubiquitous in nature. Investigating their dynamical properties is a formidable task due to the complexity of the Hilbert space they live in. Quantum computers may have the potential to effectively simulate quantum systems, provided that the many-body state under scrutiny can be faithfully prepared via an efficient algorithm. With this aim, we present a hybrid quantum--classical variational quantum algorithm for the preparation of the Gibbs state of the quantum $XY$ model. Our algorithm is based on the Grover and Rudolph parametrized quantum circuit for the preparation of the Boltzmann weights of the Gibbs state, and on a parity-preserving ansatz for the allocation of the eigenenergy basis to their respective Boltzmann weight. We explicitly show, with a paradigmatic few-body case instance, how the symmetries of a many-body system can be exploited to significantly reduce the exponentially increasing number of variational parameters needed in the Grover and Rudolph algorithm. Finally, we show that the density matrix, of the Gibbs state of the $XY$ model, obtained by statevector simulations for different parameters, exhibits a fidelity close to unity with the exact Gibbs state;  this highlights the potential use of our protocol on current quantum computers.}

\section{Introduction}
\label{sec:0}
Quantum many-body systems present formidable challenges due to the inherent complexity of the Hilbert space they encompass~\cite{Bengtsson2006}. Even the task of finding the ground-state of such systems is generally categorized as an NP-hard problem~\cite{Barahona1982}. A recent and promising approach for tackling computational problems that are computationally demanding on classical computers is to harness the power of quantum computers~\cite{Harrow2017}. However, it is important to note that the current state of quantum computation is constrained to the capabilities achievable on NISQ (Noisy Intermediate-Scale Quantum) computers~\cite{Preskill2018quantumcomputingin}. These machines are limited in that they can only apply quantum circuits of limited depth to a few noisy qubits before issues such as decoherence and gate errors compromise the accuracy of the obtained results.

In the past decade, a hybrid quantum--classical paradigm has been proposed as a means to harness the best of both classical and NISQ quantum computation~\cite{Bharti2022}. These quantum--classical algorithms delegate the classically intractable part of an algorithm to the quantum computer, while performing efficiently subroutines on a classical computer, such as objective function optimization. A paradigmatic example of such procedures are variational quantum algorithms (VQAs)~\cite{Cerezo2020}, where a parametrized quantum circuit (PQC), along with the classical optimizer, is tasked to prepare the quantum state encoding the solution to a given problem~\cite{Tilly2022}. VQAs have been applied to a variety of quantum many-body problems, ranging from determining the electronic structure of small molecules~\cite{Ratini2022}, investigating the onset of exotic phases of matter~\cite{Consiglio2022}, to detecting entanglement~\cite{Consiglio2022a}.

At the same time, one of the main applications of quantum computing is expected to be the simulation of quantum many-body systems. In particular, simulating a quantum system at thermal equilibrium is key for several disciplines, including quantum chemistry, biology and combinatorial optimisation. Evidently, preparing a quantum many-body system at thermal equilibrium is QMA-hard~\cite{Watrous2008}. Different quantum algorithms have been proposed for the preparation of thermal states based on fluctuation theorems~\cite{Holmes2022}, imaginary time evolution~\cite{Motta2020}, utilizing a generalized quantum assisted simulator~\cite{Haug2022} and via adiabatic quantum computation~\cite{Schaller2008}.

In this chapter we illustrate a thermal state preparation algorithm for a quantum many-body system based on the VQA first proposed in Ref.~\cite{Consiglio2023a}, where the PQC acts on an ancillary and a system register on which, respectively, the Boltzmann distribution and the Gibbs state are prepared. The advantage of this approach is that the von Neumann entropy, which is not an observable, can be determined without truncation. We apply our algorithm to the $XY$ spin-$\frac{1}{2}$ Heisenberg Hamiltonian utilizing the Grover and Rudolph algorithm~\cite{Grover2002} for preparing the Boltzmann distribution, highlighting how the symmetries of the model can be utilized to reduce the number of variational parameters in the circuit, followed by using a parity-preserving ansatz for diagonalizing the Hamiltonian. The chapter is organized as follows: in Sec.~\ref{sec:1} we introduce the $XY$ Hamiltonian of which we aim to prepare the thermal state; in Sec.~\ref{sec:3} we present the Grover and Rudolph algorithm used to prepare probability distribution functions on a quantum computer; in Sec.~\ref{sec:4} we implement the symmetries of the model in order to reduce the number of variational parameters used for the thermal state preparation; in Sec.~\ref{sec5} we present our VQA and its performance in preparing the $XY$ thermal state; finally, in Sec.~\ref{sec:conclusion} we draw future prospects of our approach.

\section{The model}
\label{sec:1}
In this Section we present the quantum many-body system of which we will prepare the thermal equilibrium state via a VQA. For the sake of completeness, we include a thorough derivation of its exact solution, which, due to the quadratic nature of the Hamiltonian in the fermionic representation, can be achieved both for finite and infinite system sizes. 

\begin{figure}[b]
\sidecaption
\includegraphics[width=0.5\textwidth]{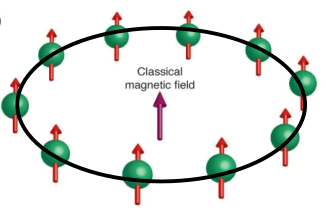}
\caption{Sketch of a 1D spin-$\frac{1}{2}$ $XY$ Heisenberg Hamiltonian with periodic boundary conditions in the presence of a magnetic field as modeled by Eq.~\eqref{eq:XY1}.}
\label{fig:1}
\end{figure}

Let us consider a 1D spin-$\frac{1}{2}$ system with periodic boundary conditions exhibiting nearest-neighbor interactions of Heisenberg type in the $XY$ plane, subject to a uniform magnetic field in the transverse $z$ direction, see Fig.~\ref{fig:1} for a finite-size instance. The $XY$ spin-$\frac{1}{2}$ Hamiltonian~\cite{LIEB1961407} is given by
\begin{align}
    \label{eq:XY}
    \hat{H}=-J\sum_{n=1}^{N}\left(\left(1+\gamma\right)\hat{S}^x_n\hat{S}^x_{n+1}+\left(1-\gamma\right)\hat{S}^y_n\hat{S}^y_{n+1}+\frac{h}{2}\hat{S}_n^z\right),
\end{align}
where $\hat{S}_i^{\alpha}$ ($\alpha=x,y,z)$ is the spin-$\frac{1}{2}$ operator sitting on site $i$ of a 1D lattice composed of $N$ even particles. The anisotropy parameter $\gamma\in\left[0,1\right]$ defines the $XX$ model ($\gamma=0$), the Ising  ($\gamma=1$) and the $XY$ model ($0<\gamma<1$), with the last two belonging to the same universality class. In the following, the diagonalization is carried out for arbitrary $\gamma$ and $h$.

One can rewrite Eq.~\eqref{eq:XY} in terms of Pauli matrices $\hat{S}_i^{\alpha}=\frac{\hat{\sigma}_i^{\alpha}}{2}$
\begin{align}
	\label{eq:XY1}
	\hat{H}=-\frac{J}{2}\sum_{n=1}^{N}\left(\frac{1+\gamma}{2}\hat{\sigma}^x_n\hat{\sigma}^x_{n+1}+\frac{1-\gamma}{2}\hat{\sigma}^y_n\hat{\sigma}^y_{n+1}+\frac{h}{2}\hat{\sigma}_n^z\right),
\end{align}
and turn to the spin ladder operators $\hat{\sigma}^{\pm}=\frac{\hat{\sigma}^x\pm i \hat{\sigma}^y}{2}$, resulting in
\begin{align}
	\label{eq:XY2}
	\hat{H}&=-\frac{J}{2}\sum_{n=1}^{N}\left(\hat{\sigma}^+_n \hat{\sigma}^-_{n+1}+\hat{\sigma}^-_n \hat{\sigma}^+_{n+1}+\gamma\left(\hat{\sigma}^+_n \hat{\sigma}^+_{n+1}+\hat{\sigma}^-_n \hat{\sigma}^-_{n+1}\right)\right)-Jh\sum_{n=1}^{N}\hat{\sigma}^+_n\hat{\sigma}^-_n+\frac{NJh}{2}\hat{\mathbb{1}}\nonumber\\
	&=-\frac{1}{2}\sum_{n=1}^{N}\left(\hat{\sigma}^+_n \hat{\sigma}^-_{n+1}+\hat{\sigma}^-_n \hat{\sigma}^+_{n+1}+\gamma\left(\hat{\sigma}^+_n \hat{\sigma}^+_{n+1}+\hat{\sigma}^-_n \hat{\sigma}^-_{n+1}\right)\right)-h\sum_{n=1}^{N}\hat{\sigma}^+_n\hat{\sigma}^-_n,
\end{align}
where in the last steps $\comm{\hat{\sigma}^+}{\hat{\sigma}^-}=\hat{\sigma}^z$ and $\acomm{\hat{\sigma}^+}{\hat{\sigma}^-}=\hat{\mathbb{1}}$ have been used. We have also set $J=1$ and rescaled the Hamiltonian by the constant term $\frac{NJh}{2}$. Applying the Jordan-Wigner transformation:
\begin{eqnarray}
	\label{eq:JW}
	\hat{c}_n=e^{-i \pi \sum_{m=1}^{n-1}\hat{\sigma}^+_m\hat{\sigma}^-_m}\hat{\sigma}^-_n,
	\hat{c}_n^{\dagger}=\hat{\sigma}^+_ne^{i \pi \sum_{m=1}^{n-1}\hat{\sigma}^+_m\hat{\sigma}^-_m},
\end{eqnarray}
the Hamiltonian in the fermionic representation reads
\begin{align}
	\label{eq:XY4}
	\hat{H}&=-\frac{1}{2}\sum_{n=1}^{N-1}\left(\hat{c}_n^{\dagger}\hat{c}_{n+1}+\gamma \hat{c}_n^{\dagger}\hat{c}_{n+1}^{\dagger}+ \text{h.c.}\right)-h\sum_{n=1}^{N}\hat{c}^{\dagger}_n\hat{c}_n -\hat{P}\left(\hat{c}_N^{\dagger}\hat{c}_1+\hat{c}_N^{\dagger}\hat{c}_1^{\dagger}+\text{h.c}\right),
\end{align}
where h.c. denotes the hermitian conjugate and the last term on the RHS accounts for the periodic boundary conditions, by defining the Parity operator $\hat{P}=\prod_{i=1}^{N}\left(-\hat{\sigma}_n^z\right)$. The Parity operator $\hat{P}$ has eigenvalues $\pm 1$, denoting, respectively, the positive and the negative parity subspace. Hence, introducing the projectors on the positive and negative parity subspaces $\hat{P}^{\pm}=\frac{1}{2}\left(1\pm \hat{P}\right)$, Eq.~\eqref{eq:XY4} can be cast in a direct sum of disjoint subspaces
\begin{align}
	\label{eq:XY5}
	\hat{H}=\hat{H}^++\hat{H}^-,~\text{where}~\hat{H}^{\pm}=\hat{P}^{\pm}\hat{H}\hat{P}^{\pm},
\end{align}
and
\begin{align}
	\label{eq:XY7}
	\hat{H}^{\pm}=-\frac{1}{2}\sum_{n=1}^{N}\left(\hat{c}_n^{\dagger}\hat{c}_{n+1}+\gamma \hat{c}_n^{\dagger}\hat{c}_{n+1}^{\dagger}+ \text{h.c.}\right)-h\sum_{n=1}^{N}\hat{c}^{\dagger}_n\hat{c}_n,
\end{align}
where antiperiodic boundary conditions, $\hat{c}_{N+1}=-\hat{c}_1$, hold for $\hat{H}^+$, and periodic boundary conditions, $\hat{c}_{N+1}=\hat{c}_1$, hold for $\hat{H}^-$. While parity effects may become relevant for finite-size systems~\cite{Damski2014,Franchini2016a}, in the thermodynamic limit $N\rightarrow \infty$, they can be neglected being of order $\mathcal{O}\left(\frac{1}{N}\right)$. In the following we will diagonalize Eq.~\eqref{eq:XY5} in the respective parity subspaces for finite size systems.

\subsection{Positive parity subspace diagonalization}

Because of the translational invariance of the system, we can perform a Fourier transformation of the fermionic operators,
\begin{align}
	\label{eq:FT}
	\hat{c}_n^{\dagger}=\frac{e^{\frac{i \pi}{4}}}{\sqrt{N}}\sum_{k\in\mathcal{K^+}}e^{ikn}\hat{c}_k^{\dagger},~\hat{c}_n=\frac{e^{\frac{-i \pi}{4}}}{\sqrt{N}}\sum_{k\in\mathcal{K^+}}e^{-ikn}\hat{c}_k,
\end{align}
where the momenta $k\in\mathcal{K^+}$ are chosen in order to satisfy the antiperiodic boundary conditions for $\hat{H}^+$, and $\mathcal{K^+}=-\frac{N-1}{N}\pi+m\frac{2\pi}{N}$, with $m=0,1,\dots,N-1$ integers. The factor $e^{\frac{i \pi}{4}}$ has been introduced so as to end up with a real Hamiltonian, reading
\begin{align}\label{eq:ham1}
\hat{H}^+=\sum_{k\in\mathcal{K^+}}\left(h-\cos k\right)~ \left(\hat{c}_k^{\dagger}\hat{c}_k+\hat{c}_{-k}^{\dagger}\hat{c}_{-k}\right)-\gamma \sin k\left(\hat{c}_k^{\dagger}\hat{c}_{-k}^{\dagger}+\hat{c}_{-k}\hat{c}_{k}\right).
\end{align}
The spinless, quadratic Hamiltonian in Eq.~\eqref{eq:ham1} is finally diagonalized via a Bogoliubov transformation
\begin{align}
   \hat{c}_k=\cos{\frac{\theta_k}{2}\hat{\gamma}_k}-\sin{\frac{\theta_k}{2}\hat{\gamma}^{\dagger}_{-k}},~\text{where}~\tan \theta_k=\frac{\gamma \sin k}{h-\cos k},
\end{align}
yielding
\begin{align}
    \label{eq:ham2}
\hat{H}^+=\sum_{k\in\mathcal{K^+}}\epsilon_k\left(\hat{\gamma}_k^{\dagger}\hat{\gamma}_k-\frac{1}{2}\right),
\end{align}
where the single-particle energy spectrum is given by
\begin{align}\label{eq_single_energy}
    \epsilon_k=\sqrt{(h-\cos k)^2 + \gamma^2\sin^2 k}.
\end{align}
The ground-state of $\hat{H}^+$ is given by the vacuum of the $\hat{\gamma}_k$ fermions and has energy
\begin{align}
    \label{eq_en_gs}
    E_0^+=-\frac{1}{2}\sum_{k\in\mathcal{K^+}}\epsilon_k,
\end{align}
and the full spectrum of $\hat{H}^+$ is derived by adding an even number of fermions to the ground-state
\begin{equation}
    E^+_k=E_0^+ + \sum_{k \in \mathcal{P}_\text{even}(\mathcal{K}^+)} \epsilon_k,
\end{equation}
where $\mathcal{P}_\text{even}(\mathcal{K}^+)$ is the subset of the power set of $\mathcal{K}^+$ with an even number of terms.

\subsection{Negative parity subspace diagonalization}

The diagonalization of the model in the negative parity subspace follows closely that in the positive one and is sketched below. In order to fulfill the periodic boundary condition for the negative parity subspace Hamiltonian $\hat{H}^-$, the Fourier transformation is carried over $k\in \mathcal{K^-}=-\pi + (m + 1) \frac{2\pi}{N}$ with $m= 0, 1, . . . , N-1$ integers. The Fourier transformation of $\hat{H}^-$ reads
\begin{align}
	\hat{H}^-&=\sum_{k\neq 0,\pi \in\mathcal{K^-}}\left(h-\cos k\right)~ \left(\hat{c}_k^{\dagger}\hat{c}_k+\hat{c}_{-k}^{\dagger}\hat{c}_{-k}\right)-\gamma \sin k\left(\hat{c}_k^{\dagger}\hat{c}_{-k}^{\dagger}+\hat{c}_{-k}\hat{c}_{k}\right)\nonumber\\
 &+\left(h-1\right)\left(\hat{c}_0^{\dagger}\hat{c}_0-\hat{c}_{0}\hat{c}_{0}^{\dagger}\right)+\left(h+1\right)\left(\hat{c}_{\pi}^{\dagger}\hat{c}_{\pi}-\hat{c}_{\pi}\hat{c}_{\pi}^{\dagger}\right),
\end{align}
where the 0-mode is occupied, whereas the $\pi$-mode is empty in the ground-state of $\hat{H}^-$. Performing a Bogoliubov rotation as in the case of $\hat{H}^+$, we obtain that the ground-state energy of $\hat{H}^-$ is given by
\begin{equation}
     E_0^-=-\frac{1}{2}\sum_{k\in\mathcal{K^-}}s(k) \epsilon_k.
\end{equation}
where
\begin{equation}
    s(k) = \begin{cases} 
      -1, & \text{if } k = 0, \\
      1, & \text{otherwise}.
   \end{cases}
\end{equation}
Similarly to $\hat{H}^+$, the full spectrum of $\hat{H}^-$ is derived by adding an even number of fermions to the ground-state
\begin{equation}
    E^-_k=E_0^- + \sum_{k \in \mathcal{P}_\text{even}(\mathcal{K}^-)} s(k) \epsilon_k,
\end{equation}
where $\mathcal{P}_\text{even}(\mathcal{K}^-)$ is the subset of the power set of $\mathcal{K}^-$ with an even number of terms.

\subsection{Gibbs density matrix}

A system, with Hamiltonian $\hat{H}$, at thermal equilibrium with a bath at (inverse) temperature $\beta=\frac{1}{k_\text{B} T}$, where $k_\text{B}$ is the Boltzmann constant, is described by a density matrix $\hat{\rho}(\beta, \hat{H})$, dubbed the Gibbs state: $\hat{\rho}(\beta, \hat{H})=\frac{e^{-\beta \hat{H}}}{\mathcal{Z}(\beta, \hat{H})}$, where $\mathcal{Z(\beta, \hat{H})}=\Tr{e^{-\beta \hat{H}}}$ is the partition function.

Because of the direct sum structure of the Hamiltonian in Eq.~\eqref{eq:XY5}, the thermal state of the spin system can also be cast in a direct sum form:
\begin{align}
    \hat{\rho}(\beta, \hat{H})=\sum_{n=0}^{2^{N-1} - 1} \left(\frac{e^{-\beta E^+_n}}{\mathcal{Z}(\beta, \hat{H})}\ketbra{E_n^+}+\frac{e^{-\beta E^-_n}}{\mathcal{Z}(\beta, \hat{H})}\ketbra{E_n^-}\right),
\end{align}
where $\ket{E_n^{\pm}}$ are the eigenstates of $\hat{H}^{\pm}$ and $p_n=\frac{e^{-\beta E^{\pm}_n}}{\mathcal{Z}(\beta, \hat{H})}$ are the Boltzmann weights giving the probability distribution function (PDF) of the Gibbs distribution. The latter is composed by
\begin{align}\label{eq:pdfG}
    \left\{p_n\right\}=\left\{p_n^+\right\}\cup\left\{p_n^-\right\}.
\end{align}
From the non-interacting nature of the Hamiltonian and its set of symmetries, it is evident that there is a high degree of degeneracy in the Hamiltonian spectrum. In the following we will analytically determine how many distinct $p_n$'s appear in Eq.~\eqref{eq:pdfG}, in order to minimize the number of necessary variational parameters in the PQC, capable of reproducing the Boltzmann distribution, and hence the Gibbs state.

\subsection{Degeneracies in the energy spectrum}\label{sS:deg}

Let us start identifying the number of distinct energy levels in the positive parity subspace. Because of the translational symmetry, the single-particle energy spectrum is invariant under momentum inversion $k\rightarrow -k$, as evident from Eq.~\eqref{eq_single_energy} being an even function of $k$: $\epsilon_k=\epsilon_{-k}$.

Clearly, the state with zero fermions, i.e., the ground-state, is non-degenerate. The next set of energies we analyze is that with two fermions added to the ground-state, that is $E_n^{\left(2\right)}=E_0^++\epsilon_k+\epsilon_q$, where $k\neq q\in \mathcal{K}^+$.
The number of 2-fermions energy levels is given by the binomial $\binom{N}{2}$ of which $\frac{N}{2}$
energy levels are non-degenerate. This can be readily seen, as degenerate levels occur in the form 
\begin{align}\label{eq_2deg}
    \epsilon_{-k}+\epsilon_{-q}=\epsilon_{k}+\epsilon_{q}=\epsilon_{k}+\epsilon_{-q}=\epsilon_{-k}+\epsilon_{q},
\end{align}
with $k \neq q$. On the other hand, non-degenerate levels occur in the form
\begin{align}\label{eq_2nondeg}
    \epsilon_{k}+\epsilon_{-k}.
\end{align}
Hence, non-degenerate levels in the two-fermions excitation spectrum are given by how many distinct $k$'s can be selected in Eq.~\eqref{eq_2nondeg}. This number is obtained by observing that the momenta $k\in\mathcal{K^+}$ are symmetric around 0, and hence the number of ways to chose one momentum $k$ out of $\frac{N}{2}$ momenta $k$ is trivially given by $\frac{N}{2}$. Moreover, as Eqs.~\eqref{eq_2deg} give rise to the same energy, the degeneracy is always four-fold in this particle sector. To summarise, out of the $\binom{N}{2}$ energy levels, there are $\binom{\frac{N}{2}}{2}$ 4-fold degenerate energy levels, with the remaining $\frac{N}{2}$ being non-degenerate.

Moving to the 4-fermion excited states, one can proceed in a similar way by identifying, out of the $\binom{N}{4}$ 4-fermion excitation energy levels, as non-degenerate levels being those with energy
\begin{align}\label{eq_4nondeg}
    &\epsilon_{-k}+\epsilon_{k}+\epsilon_{-q}+\epsilon_{q},
\end{align}
where $k\neq q$. Hence, the non-degenerate levels of the 4-fermion excited states are $\binom{\frac{N}{2}}{2}$, corresponding to the number of distinct sets $\left(k,q\right)$ that can be selected from the available momenta in the positive subset of $\mathcal{K^+}$. The degrees of degeneracy in the 4-particles sector are $\left(1,4,16\right)$ occurring, respectively,  $\left(\binom{\frac{N}{2}}{2},\binom{\frac{N}{2}}{1} \binom{\frac{N}{2}-1}{2},\binom{\frac{N}{2}}{4}\right)$ times.

Generalizing the above procedure to an arbitrary $n$-fermion energy subspace, for a chain of length $N$, one obtains that out of the $\binom{N}{n}$ $n$-fermion excitation energy levels
\begin{itemize}
    \item the number of 1-fold degenerate levels is $\binom{\frac{N}{2}}{\frac{n}{2}}$
    \item the number of 4-fold degenerate levels is $\binom{\frac{N}{2}}{\frac{n}{2}-1}\binom{\frac{N}{2}-\left(\frac{n}{2}-1\right)}{2}$
    \item the number of 16-fold degenerate levels is $\binom{\frac{N}{2}}{\frac{n}{2}-2}\binom{\frac{N}{2}-\left(\frac{n}{2}-2\right)}{4}$
    \item the number of 64-fold degenerate levels is $\binom{\frac{N}{2}}{\frac{n}{2}-3}\binom{\frac{N}{2}-\left(\frac{n}{2}-3\right)}{6}$
    \item the number of $4^j$-fold degenerate levels is $\binom{\frac{N}{2}}{\frac{n}{2}-j}\binom{\frac{N}{2}-\left(\frac{n}{2}-j\right)}{2j}$
\end{itemize}
Clearly, the maximum degree of degeneracy in the $n$-fermion energy subspace is $4^j \leq \binom{N}{n}$, with $j$ being a non-negative integer. The procedure just described determines the degree of degeneracy and the number of energy levels in each degenerate subspace of $\hat{H}^+$ can be straightforwardly applied also to the negative parity subspace, which will not be reported here to avoid redundancy.

\section{Preparing probability distributions on a quantum computer}\label{sec:3}

In this Section we move to the main aim of this chapter: preparing a thermal state via a VQA. First we focus on preparing the Boltzmann distribution of the $XY$ model entailed by Eq.~\eqref{eq:pdfG}.

Grover and Rudolph~\cite{Grover2002} first described a procedure to prepare log-concave probability density functions on a quantum computer. However, this approach was later proven that it does not achieve a quantum speed-up in general~\cite{Herbert2021}. Modern approaches resort to VQAs~\cite{Nakaji2022, Dasgupta2022, Kumar2023}, whereby parametrized ansätze are used to prepare the probability distribution on a quantum register: by varying the parameters and minimizing a relevant cost function, such as the relative entropy. Similar, in this vein, is the generative adversarial network (GAN), where the classical optimizer is replaced by a classical neural network, acting as a discriminator for the probability distribution~\cite{Zoufal2019, Situ2020, Agliardi2022}.

\begin{figure}[t]
    \centering
    \includegraphics[width=\textwidth]{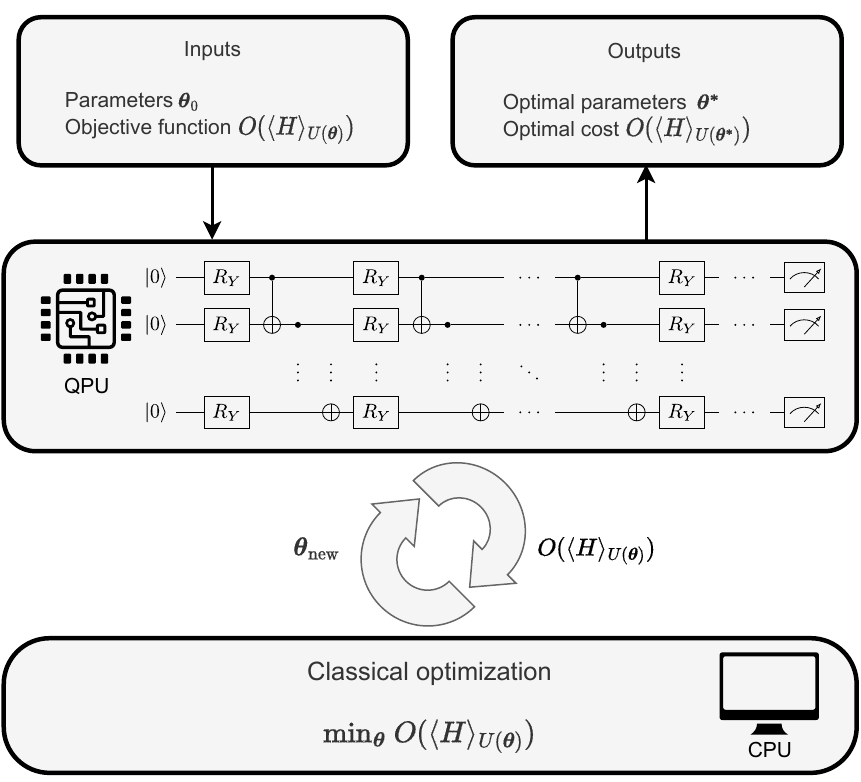}
    \caption{Diagrammatic representation of a VQA. The framework of a VQA requires: an objective function (inputs), a parameterized quantum circuit (that operates on the quantum processing unit); and a classical optimizer (that operates on the classical processing unit). The Inputs are fed into the QPU (upper left arrow), the output of the PQC is fed into the CPU, which determines new parameters $\bm{\theta}_{\text{new}}$ provided to the QPU for the next iteration. This procedure is looped until optimal parameters $\bm{\theta^*}$ are found, which constitute the Outputs (upper right arrow).}
    \label{fig:vqa}
\end{figure}

A VQA comprises several modular components that can be readily combined, extended, and improved with developments in quantum hardware and theory~\cite{Bharti2022}. These components include: the objective function, the cost function to be variationally optimized; the PQC, composed of unitary gates that are updated in the optimization procedure; and the classical optimizer, the method used to obtain the optimal circuit parameters that optimize the objective function. Fig.~\ref{fig:vqa} highlights the modules of a VQA.

To prepare the Boltzmann distribution of the $XY$ model, we take the approach of utilizing a VQA, alongside a parametrized ansatz in the form of Grover and Rudolph~\cite{Grover2002}. In this case, an $N$-qubit ansatz requires $2^N - 1$ parameters, and $2^N - 1$ parameterized $R_y$ gates, most of which are multi-controlled. An $R_y(\theta)$ gate acting on one qubit is given by
\begin{equation}
    R_y(\theta) = \left( \begin{array}{c c} \cos{\frac{\theta}{2}} & -\sin{\frac{\theta}{2}} \\[4pt] \sin{\frac{\theta}{2}} & \cos{\frac{\theta}{2}}\end{array} \right).
\end{equation}
Controlled gates act on two or more qubits, where one or more qubits act as a control for some operation. The most common example is the $\textsc{CNOT}$ gate, which performs the $\textsc{NOT}$ operation on the second qubit only when the first qubit is the state $\ket{1}$, otherwise it leaves the state unchanged. In quantum circuit diagrams, a black dot represents the control qubit activating the controlled gate when in the state $\ket{1}$, while a white dot represents the control qubit activating the controlled gate when in the state $\ket{0}$. As a result, since the Grover and Rudolph algorithm serves to create a probability distribution from the entries of the first column, it utilizes only parametrized orthogonal (real unitary) gates. Note that an extended algorithm was proposed by Ref.~\cite{Benenti2009} to obtain an optimal purification of an $n$-qudit state.

\begin{figure}[t]
    \centering
    \includegraphics[width=\textwidth]{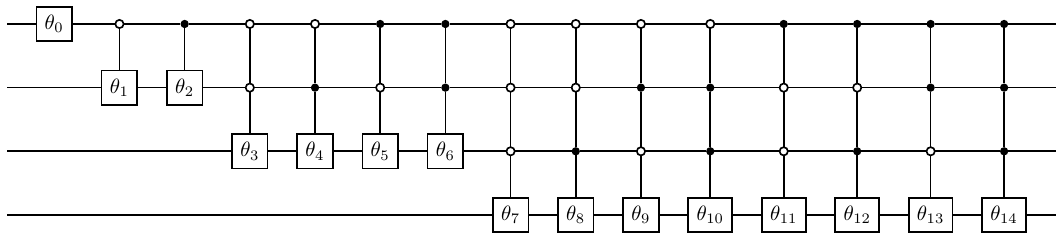}
    \caption{Grover and Rudolph Circuit for preparing an arbitrary probability distribution embedded in a real quantum state. Each gate is represented an $R_y$ gate with its respective parameter $\theta_i$. Black dots represent the control qubit activating on $\ket{1}$, while the white dots represent the control qubit activating on $\ket{0}$.}
    \label{fig:circuit}
\end{figure}

Let us consider the case for $N = 4$, with Fig.~\ref{fig:circuit} showing the circuit. Each box represents the $R_y$ gate with the relevant parameter, and the black and white dots represent the control qubits. For example, the gate with $R_y(\theta_9)$ acts only on qubit 3 if qubit 0 is in the state $\ket{0}$, qubit 1 is in the state $\ket{1}$, and qubit 2 is in the state $\ket{0}$. Since there are $2^k$ possible binary controls for a $k$-qubit controlled gate, one can see how the circuit scales exponentially, both in number of gates and number of parameters. Our goal in this manuscript is the reduce the number of parameters in the circuit shown in Fig.~\ref{fig:circuit}, by exploiting the symmetries of the $XY$ model giving rise to the degeneracies in the energy levels obtained in Sec.~\ref{sS:deg}.

The action of the circuit $U$ in Fig.~\ref{fig:circuit} acting on the all-zero four-qubit state results in
\begin{equation}
    \label{eq:U}
    U\ket{0000} = \sum_{i=0000}^{1111} a_i \ket{i}.
\end{equation}
The states $\ket{0}$ and $\ket{1}$ are the computational basis states of one qubit. As such, extending the computational basis states for $n$ qubits is simply done by computing the tensor product of all combinations of the states $\ket{0}$ and $\ket{1}$ up to $n$ qubits. For example, the computational basis states of two qubits are $\ket{00}, \ket{01}, \ket{10}, \ket{11}$. As a result, when measuring the quantum state in Eq.~\eqref{eq:U} in the computational basis, we end up with the following probabilities for each configuration:
{\allowdisplaybreaks
\begin{subequations}\label{eq:ration}
\begin{align}
    \cos^2\left(\frac{\theta_0}{2}\right) \cos^2\left(\frac{\theta_1}{2}\right) \cos^2\left(\frac{\theta_3}{2}\right) \cos^2\left(\frac{\theta_7}{2}\right) &= p_{0000}, \\
    \cos^2\left(\frac{\theta_0}{2}\right) \cos^2\left(\frac{\theta_1}{2}\right) \cos^2\left(\frac{\theta_3}{2}\right) \sin^2\left(\frac{\theta_7}{2}\right) &= p_{0001}, \\
    \cos^2\left(\frac{\theta_0}{2}\right) \cos^2\left(\frac{\theta_1}{2}\right) \sin^2\left(\frac{\theta_3}{2}\right) \cos^2\left(\frac{\theta_8}{2}\right) &= p_{0010}, \\
    \cos^2\left(\frac{\theta_0}{2}\right) \cos^2\left(\frac{\theta_1}{2}\right) \sin^2\left(\frac{\theta_3}{2}\right) \sin^2\left(\frac{\theta_8}{2}\right) &= p_{0011}, \\
    \cos^2\left(\frac{\theta_0}{2}\right) \sin^2\left(\frac{\theta_1}{2}\right) \cos^2\left(\frac{\theta_4}{2}\right) \cos^2\left(\frac{\theta_9}{2}\right) &= p_{0100}, \\
    \cos^2\left(\frac{\theta_0}{2}\right) \sin^2\left(\frac{\theta_1}{2}\right) \cos^2\left(\frac{\theta_4}{2}\right) \sin^2\left(\frac{\theta_9}{2}\right) &= p_{0101}, \\
    \cos^2\left(\frac{\theta_0}{2}\right) \sin^2\left(\frac{\theta_1}{2}\right) \sin^2\left(\frac{\theta_4}{2}\right) \cos^2\left(\frac{\theta_{10}}{2}\right) &= p_{0110}, \\
    \cos^2\left(\frac{\theta_0}{2}\right) \sin^2\left(\frac{\theta_1}{2}\right) \sin^2\left(\frac{\theta_4}{2}\right) \sin^2\left(\frac{\theta_{10}}{2}\right) &= p_{0111}, \\
    \sin^2\left(\frac{\theta_0}{2}\right) \cos^2\left(\frac{\theta_2}{2}\right) \cos^2\left(\frac{\theta_5}{2}\right) \cos^2\left(\frac{\theta_{11}}{2}\right) &= p_{1000}, \\
    \sin^2\left(\frac{\theta_0}{2}\right) \cos^2\left(\frac{\theta_2}{2}\right) \cos^2\left(\frac{\theta_5}{2}\right) \sin^2\left(\frac{\theta_{11}}{2}\right) &= p_{1001}, \\
    \sin^2\left(\frac{\theta_0}{2}\right) \cos^2\left(\frac{\theta_2}{2}\right) \sin^2\left(\frac{\theta_5}{2}\right) \cos^2\left(\frac{\theta_{12}}{2}\right) &= p_{1010}, \\
    \sin^2\left(\frac{\theta_0}{2}\right) \cos^2\left(\frac{\theta_2}{2}\right) \sin^2\left(\frac{\theta_5}{2}\right) \sin^2\left(\frac{\theta_{12}}{2}\right) &= p_{1011}, \\
    \sin^2\left(\frac{\theta_0}{2}\right) \sin^2\left(\frac{\theta_2}{2}\right) \cos^2\left(\frac{\theta_6}{2}\right) \cos^2\left(\frac{\theta_{13}}{2}\right) &= p_{1100}, \\
    \sin^2\left(\frac{\theta_0}{2}\right) \sin^2\left(\frac{\theta_2}{2}\right) \cos^2\left(\frac{\theta_6}{2}\right) \sin^2\left(\frac{\theta_{13}}{2}\right) &= p_{1101},\\
    \sin^2\left(\frac{\theta_0}{2}\right) \sin^2\left(\frac{\theta_2}{2}\right) \sin^2\left(\frac{\theta_6}{2}\right) \cos^2\left(\frac{\theta_{14}}{2}\right) &= p_{1110}, \\
    \sin^2\left(\frac{\theta_0}{2}\right) \sin^2\left(\frac{\theta_2}{2}\right) \sin^2\left(\frac{\theta_6}{2}\right) \sin^2\left(\frac{\theta_{14}}{2}\right) &= p_{1111}.
\end{align}
\end{subequations}
}
If one knows beforehand the discrete probability distribution, then, by taking specific ratios of expressions in Eqs.~\eqref{eq:ration}, one can relate the parameters $\theta_i$ to ratios of probabilities, e.g., 
\begin{align}
    \frac{p_{1111}}{p_{1110}}=\tan^2\left(\frac{\theta_{14}}{2}\right)\implies \theta_{14}=2 \arctan\left( \sqrt{\frac{p_{1111}}{p_{1110}}}\right).
\end{align}
By iterating this procedure for all $\theta$, one can readily obtain
\begin{equation}
    \theta_{l + 2^{N - k} - 1} = 2\arctan\left(\sqrt{\frac{\sum_{i=0}^{k-1} p_{2^k l + i + k}}{\sum_{j=0}^{k-1} p_{2^k l + j}}}\right),
\end{equation}
where $k$ ranges from $1, \dots, N$, and $l$  from $0, \dots, 2^{N-k} - 1$.

\section{Few-body \textit{XY} chain}

\label{sec:4}
To illustrate an instance of the Gibbs PDF preparation via the Grover and Rudolph algorithm where the degeneracies of the energy levels allow for a reduction of the number of variational parameters needed in the PQC, let us restrict to a few-body scenario. This allows to better keep track of the protocol relying on the analytic expressions of the eigenenergies.  Let us focus on the $N=4$ $XY$ chain.

\begin{figure}[t]
    \centering
    \includegraphics[width=\textwidth]{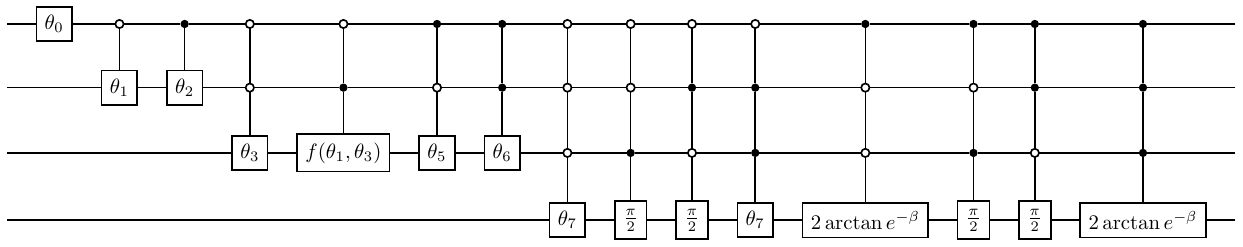}
    \caption{Fig.~\ref{fig:circuit} with substituted parameters, Eqs.~\eqref{eq:first}~-~\eqref{eq:last}, where $f(\theta_1, \theta_3) = 2\arccos\left(\frac{\sin\left(\frac{\theta_{3}}{2}\right)}{\tan\left(\frac{\theta_{1}}{2}\right)}\right)$.}
    \label{fig:circuit_substituted}
\end{figure}

Since the $XY$ model is non-interacting, the spectrum is given by the single-particle energies, as discussed in Section~\ref{sec:1}. The model is also symmetric, when the eigenenergies are calculated with the modes ordered lexicographically, we can assign the first half of the computational basis states to cover $\left\{p_n^+\right\}$, with the latter half to cover $\left\{p_n^-\right\}$. The eigenenergies are:
\begin{subequations}
\begin{align}
      E_0^+ &= \frac{-\sqrt{\gamma ^2+2 h \left(h-\sqrt{2}\right)+1}-\sqrt{\gamma ^2+2 h
   \left(h+\sqrt{2}\right)+1}}{\sqrt{2}}, \\
    E_1^+ &= \frac{\sqrt{\gamma ^2+2 h \left(h-\sqrt{2}\right)+1}-\sqrt{\gamma ^2+2 h
	   \left(h+\sqrt{2}\right)+1}}{\sqrt{2}}, \\
    E_2^+ &= E_3^+ = E_4^+ = E_5^+ = 0, \\
    E_6^+ &= \frac{-\sqrt{\gamma ^2+2 h
   \left(h-\sqrt{2}\right)+1}+\sqrt{\gamma ^2+2 h \left(h+\sqrt{2}\right)+1}}{\sqrt{2}}, \\
    E_7^+ &= \frac{\sqrt{\gamma ^2+2 h \left(h-\sqrt{2}\right)+1}+\sqrt{\gamma ^2+2 h
   \left(h+\sqrt{2}\right)+1}}{\sqrt{2}}, \\
    E_0^- &= -1-\sqrt{\gamma ^2+h^2},~
    E_1^- = 1-\sqrt{\gamma ^2+h^2}, \\
    E_2^- &= E_3^- = -E_4^- = -E_5^- = -h, \\
    E_6^- &= -1+\sqrt{\gamma ^2+h^2},~
    E_7^- = 1+\sqrt{\gamma ^2+h^2}.
\end{align}
\end{subequations}
We start to look at the Boltzmann coefficients generated by the eigenenergies in the negative parity sector. In the case of the 
\begin{align}
    \frac{p_{1111}}{p_{1110}} &= \tan^2\left(\frac{\theta_{14}}{2}\right) = e^{-\beta\left(E_{7}^- - E_{6}^-\right)} = e^{-2\beta} \implies \theta_{14} = 2\arctan\left(e^{-\beta}\right).
    \label{eq:first}
\end{align}
Next we find that
\begin{align}
    \frac{p_{1101}}{p_{1100}} &= \tan^2\left(\frac{\theta_{13}}{2}\right) = e^{-\beta\left(E_{5}^- - E_{4}^-\right)} = 1 \implies \theta_{13} = \frac{\pi}{2},
\end{align}
and
\begin{align}
    \frac{p_{1011}}{p_{1010}} &= \tan^2\left(\frac{\theta_{12}}{2}\right) = e^{-\beta\left(E_{3}^- - E_{2}^-\right)} = 1 \implies \theta_{12} = \frac{\pi}{2},
\end{align}
due to the pairs of degeneracies. Furthermore
\begin{align}
    \frac{p_{1001}}{p_{1000}} = \tan^2\left(\frac{\theta_{11}}{2}\right) = e^{-\beta\left(E_{1}^- - E_{0}^-\right)} = e^{-2\beta} \implies \theta_{11} = 2\arctan\left(e^{-\beta}\right).
\end{align}
Now looking at the Boltzmann coefficients given by the eigenenergies in the positive parity sector:
\begin{align}
    \frac{p_{0101}}{p_{0100}} = \tan^2\left(\frac{\theta_{9}}{2}\right) = e^{-\beta\left(E_{5}^+ - E_{4}^+\right)} = 1 \implies \theta_{9} = \frac{\pi}{2},
\end{align}
and
\begin{align}
    \frac{p_{0011}}{p_{0010}} = \tan^2\left(\frac{\theta_{8}}{2}\right) = e^{-\beta\left(E_{3}^+ - E_{2}^+\right)} = 1 \implies \theta_{8} = \frac{\pi}{2}.
\end{align}
Moreover, while we cannot determine the exact value of $\theta_{10}$ and $\theta_7$, due to their dependency on $\gamma$ and $h$, we can still relate them through
\begin{align}
    \frac{p_{0111}}{p_{0110}}= \frac{p_{0001}}{p_{0000}} 
    \implies \tan^2\left(\frac{\theta_{10}}{2}\right)= \tan^2\left(\frac{\theta_7}{2}\right) 
    \implies \theta_{10}= \theta_7.
\end{align}
Finally, we also have that
\begin{align}
    \frac{p_{0101}}{p_{0011}} &= \frac{\cos^2\left(\frac{\theta_{4}}{2}\right)}{\sin^2\left(\frac{\theta_{3}}{2}\right)} \tan^2\left(\frac{\theta_{1}}{2}\right) = e^{-\beta\left(E_{5}^+ - E_{3}^+\right)} = 1 \nonumber \\
    &\implies \theta_{4} = 2\arccos\left(\frac{\sin\left(\frac{\theta_{3}}{2}\right)}{\tan\left(\frac{\theta_{1}}{2}\right)}\right).
    \label{eq:last}
\end{align}
Thus we have managed to eliminate 8 parameters from a total of 15. This may be due to the fact that we have $N$ single-particle energy levels in both the positive and the negative parity subspace. However we also have the fact that in the negative parity subspace, the sum of the energies of the $\pi$- and 0-modes sum always up to 2, hence giving an additional constraint. This result may hint towards the fact that for the $XY$ model composed of $N$ particles, the number of parameters needed to characterize its Boltzmann distribution is $2N - 1$. Fig.~\ref{fig:circuit_substituted} represents the PQC in Fig.~\ref{fig:circuit} with the reduced number of parameters obtained from Eqs.~\eqref{eq:first}~-~\eqref{eq:last}.

\section{Structure of the VQA} \label{sec5}

We utilize the algorithm proposed in Ref.~\cite{Consiglio2023a} to prepare Gibbs states of the $XY$ model. Suppose one chooses a Hamiltonian $\hat{H}$ and inverse temperature $\beta$. For a general state $\hat{\rho}$, one can define a generalized Helmholtz free energy as
\begin{equation}
    \mathcal{F}(\hat{\rho}) = \Tr{\hat{H} \hat{\rho}} - \beta^{-1}\mathcal{S}(\hat{\rho}),
    \label{eq:helmholtz_free_energy}
\end{equation}
where the von Neumann entropy $\mathcal{S}(\hat{\rho})$ can be expressed in terms of the eigenvalues, $p_i$, of $\hat{\rho}$,
\begin{equation}
    \mathcal{S}(\hat{\rho}) = -\sum_{i=0}^{d - 1} p_i \ln p_i.
    \label{eq:von_neumann}
\end{equation}
Since the Gibbs state is the unique state that minimizes the free energy, a variational form can be put forward that takes Eq.~\eqref{eq:helmholtz_free_energy} as an objective function, such that
\begin{equation}
    \hat{\rho}(\beta, \hat{H}) = \underset{\hat{\rho}}{\arg\min}~\mathcal{F}(\hat{\rho}).
\end{equation}
In this case, $p_i = \exp\left(-\beta E_i\right)/\mathcal{Z}(\beta, \hat{H})$ is the probability of getting the eigenstate $\ket{E_i}$ from the ensemble $\hat{\rho}(\beta, \hat{H})$.

\begin{figure}[t]
    \centering
    \includegraphics[width=0.5\textwidth]{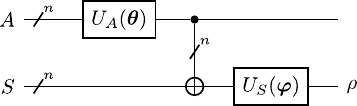}
    \caption{PQC for Gibbs state preparation, with systems $A$ and $S$ each carrying $N$ qubits. \textsc{CNOT} gates act between each qubit $A_i$ and corresponding $S_i$.}
    \label{fig:gibbs_circuit}
\end{figure}

The PQC, as shown in Fig.~\ref{fig:gibbs_circuit}, is composed of a unitary gate $U_A$ acting on the ancillary qubits, and a unitary gate $U_S$ acting on the system qubits, with \textsc{CNOT} gates in between. Note that the circuit notation we are using here means that there are $N$ qubits for both the system and the ancillae, as well as $N$ \textsc{CNOT} gates that act in parallel, $\textsc{CNOT}_{AS} \equiv \bigotimes_{i=0}^{N - 1}\textsc{CNOT}_{A_i S_i}$. The parametrized unitary $U_A$ acting on the ancillae, followed by \textsc{CNOT} gates between the ancillary and system qubits, is responsible for preparing a probability distribution on the system. The parametrized unitary $U_S$ is then applied on the system qubits to transform the computational basis states into the eigenstates of the Hamiltonian.

Throughout the paper we will describe unitary operations and density matrices in the computational basis. We will also denote the $N$-fold tensor product of a state $\ket{\psi}$, of an $N$-qubit register $R$, as $\ket{\psi}_R^{\otimes N} \equiv \bigotimes_{i = 0}^{N - 1} \ket{\psi}_i$, where $i \in R$. Denote a general unitary gate of dimension $d = 2^N$, as $U_A = (u_{i, j})_{0\leq i, j \leq d-1}$. Starting with the initial state of the $2n$-qubit register, $\ket{0}_{AS}^{\otimes 2n}$, we apply the unitary gate $U_A$ on the ancillae to get a quantum state $\ket{\psi}_A$, such that $(U_A \otimes \hat{\mathbb{1}}_S) \ket{0}_{AS}^{\otimes 2n} = \ket{\psi}_{A} \otimes \ket{0}_{S}^{\otimes N}$, where $\ket{\psi}_A = \sum_{i=0}^{d - 1} u_{i,0} \ket{i}_A$ and $\mathbb{1}_S$ is the identity acting on the system. Since $U_A$ is applied to the all-zero state in the ancillary register, then the operation serves to extract the first column of the unitary operator $U_A$, as discussed in Section~\ref{sec:3}. The next step is to prepare a classical probability mixture on the system qubits, which can be done by applying \textsc{CNOT} gates between each ancilla and system qubit. This results in a state
\begin{align}
    \textsc{CNOT}_{AS} \left( \ket{\psi}_{A} \otimes \ket{0}_{S}^{\otimes N} \right) = \sum_{i=0}^{d - 1} u_{i,0} \ket{i}_A \otimes \ket{i}_S.
\end{align}
By then tracing out the ancillary qubits, we arrive at
\begin{align}
    \PTr{A}{\left( \sum_{i=0}^{d-1} u_{i,0}\ket{i}_A \otimes \ket{i}_S \right) \left( \sum_{j=0}^{d-1} u_{j,0}^* \bra{j}_A \otimes \bra{j}_S \right)}= \sum_{i=0}^{d-1} |u_{i,0}|^2 \ketbra{i}{i}_S,
    \label{eq:diag}
\end{align}
ending up with a diagonal mixed state on the system, with probabilities given directly by the absolute square of the entries of the first column of $U_A$, that is, $p_i = |u_{i,0}|^2$. If the system qubits were traced out instead, we would end up with the same diagonal mixed state,
\begin{align}
    \PTr{S}{\left( \sum_{i=0}^{d-1} u_{i,0}\ket{i}_A \otimes \ket{i}_S \right) \left( \sum_{j=0}^{d-1} u_{j,0}^* \bra{j}_A \otimes \bra{j}_S \right)} =
    \sum_{i=0}^{d-1} |u_{i,0}|^2 \ketbra{i}{i}_A.
\end{align}
This implies that by measuring in the computational basis of the ancillary qubits, we can determine the probabilities $p_i$, which can then be post-processed to determine the von Neumann entropy $\mathcal{S}$ of the state $\hat{\rho}$ via Eq.~\eqref{eq:von_neumann} (since the entropy of $A$ is the same as that of $S$). 

\begin{figure}[t]
    \centering
    \includegraphics[width=\textwidth]{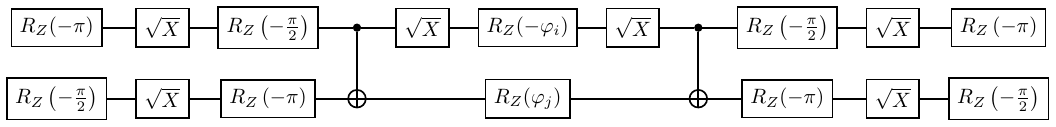}
    \caption{Decomposed $R_P$ gate in Eq.~\eqref{eq:R_P} in IBM native basis gates.}
    \label{fig:R_P}
\end{figure}

\begin{figure}[t]
    \centering
    \includegraphics[width=0.8\textwidth]{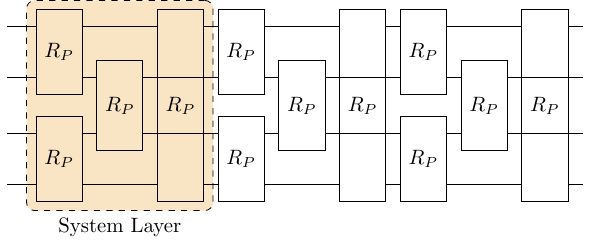}
    \caption{A representation of the PQC embodying $U_S$ tasked with performing the rotation form the computational to the eigenenergy basis of the system Hamiltonian. The shaded area represent a single brick-wall layer that is repeated three times in obtaining the fidelity of Fig.~\ref{fig:fidelity_plot}. Note that each $R_P$ gate contains 2 parameters as in Eq.~\eqref{eq:R_P}.}
    \label{fig:U_S}
\end{figure}

The unitary gate $U_S$ then serves to transform the computational basis states of the system qubits to the eigenstates of the Gibbs state, once it is optimized by the VQA, such that
\begin{align}
    \hat{\rho}= U_S \left( \sum_{i=0}^{d-1} |u_{i,0}|^2 \ketbra{i}{i}_S \right) U_S^\dagger= \sum_{i=0}^{d - 1} p_i \ketbra{\psi_i},
\end{align}
where the expectation value $\Tr{\hat{H} \hat{\rho}}$ of the Hamiltonian can be measured. Ideally, at the end of the optimization procedure, $p_i = \exp\left(-\beta E_i\right)/\mathcal{Z}(\beta, \hat{H})$ and $\ket{\psi_i} = \ket{E_i}$, so that we get
\begin{equation}
    \hat{\rho}(\beta, \hat{H}) = \sum_{i=0}^{d - 1} \frac{e^{-\beta E_i}}{\mathcal{Z}(\beta, \hat{H})} \ketbra{E_i}.
\end{equation}
Finally, we define the objective function of our VQA to minimize the free energy~\eqref{eq:helmholtz_free_energy}, via our constructed PQC, to obtain the Gibbs state
\begin{align}
    \hat{\rho}(\beta, \hat{H}) &= \underset{\bm{\theta}, \bm{\varphi}}{\arg\min} \ \mathcal{F}\left(\hat{\rho}\left(\bm{\theta}, \bm{\varphi}\right)\right) \nonumber \\ 
    &= \underset{\bm{\theta}, \bm{\varphi}}{\arg\min} \left( \Tr{\hat{H} \hat{\rho}_S(\bm{\theta}, \bm{\varphi})} - \beta^{-1}\mathcal{S}\left(\hat{\rho}_A(\bm{\theta})\right) \right).
    \label{eq:gibbs_state}
\end{align}
After optimization, the system register will (ideally) contain the prepared Gibbs state. In Refs.~\cite{Consiglio2023a, Consiglio2023b}, $U_A$ is the ansatz responsible for preparing the Boltzmann distribution, while $U_S$ is the ansatz responsible for preparing the eigenstates of the Hamiltonian. In this work, we replace $U_A$ in Fig.~\ref{fig:gibbs_circuit} by the ansatz in Fig.~\ref{fig:circuit_substituted}, and optimize to obtain the Gibbs state of the $XY$ model. $U_S$ is a parity-preserving ansatz, given by a brick-wall layering of $R_P$ gates, denoted as
\begin{align}
    R_P(\varphi_i, \varphi_j) = \left( \begin{array}{cccc}
     \cos \left(\frac{\varphi_i +\varphi_j }{2}\right) & 0 & 0 & \sin \left(\frac{\varphi_i +\varphi_j }{2}\right) \\
     0 & \cos \left(\frac{\varphi_i -\varphi_j }{2}\right) & -\sin \left(\frac{\varphi_i -\varphi_j }{2}\right) & 0 \\
     0 & \sin \left(\frac{\varphi_i -\varphi_j }{2}\right) & \cos \left(\frac{\varphi_i -\varphi_j }{2}\right) & 0 \\
     -\sin \left(\frac{\varphi_i +\varphi_j }{2}\right) & 0 & 0 & \cos \left(\frac{\varphi_i +\varphi_j }{2}\right) \\
    \end{array}
    \right).
    \label{eq:R_P}
\end{align}
As one can see from the unitary itself, any input state will have the same parity as the output state, and so, an ansatz constructed solely of parity-preserving gates will also be parity-preserving. The decomposed unitary gate is shown in Fig.~\ref{fig:R_P}, while the entire circuit of $U_S$ is shown in Fig.~\ref{fig:U_S}.

\begin{figure}[t]
    \centering
    \includegraphics[width=\textwidth]{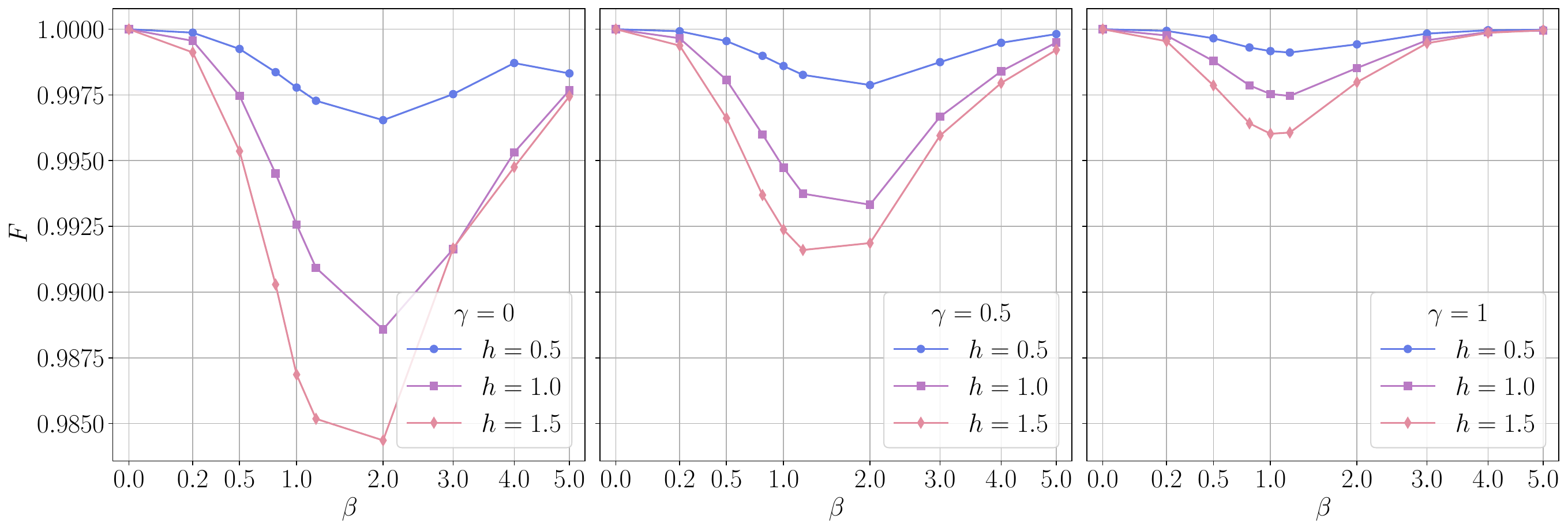}
    \caption{Fidelity $F$ between the obtained statevector simulations (using BFGS) and the exact Gibbs state, vs inverse temperature $\beta$, for the $XY$ model with $h = 0.5, 1, 1.5$ and $\gamma=0, 0.5, 1$ in Eq.~\eqref{eq:XY1} for four qubits. A total of 100 runs are made for each point, with the optimal state taken to be the one that maximizes the fidelity.}
    \label{fig:fidelity_plot}
\end{figure}

To assess the performance of the VQA, we utilize the Uhlmann-Josza fidelity as a figure of merit~\cite{Uhlmann2011}, defined as $F\left(\hat{\rho},\hat{\sigma}\right) = \left(\Tr{\sqrt{\sqrt{\hat{\rho}}\hat{\sigma}\sqrt{\hat{\rho}}}}\right)^2$. This fidelity measure quantifies the ``closeness'' between the prepared state and the target Gibbs state, making it a commonly-employed metric for distinguishability. Fig.~\ref{fig:fidelity_plot} shows the fidelity of the prepared Gibbs state with the theoretical Gibbs state of the four-qubit $XY$ model at $h = 0.5, 1, 1.5$ and $\gamma=0, 0.5, 1$, across a broad range of temperatures. Three layers of both $U_A$ and $U_S$ were used for all simulations. Fig.~\ref{fig:fidelity_plot} shows that the fidelity, while dipping at intermediary temperatures, still achieves values greater than 98\% for all models and temperatures. Interestingly, the results for $\gamma=0$ appear to be slightly less optimal. This may due to the fact that the $XX$ model belongs to a different universality class than the $XY$ and Ising model, exhibiting an additional symmetry, the conservation of the total magnetization.

\section{Conclusion} \label{sec:conclusion}

In this chapter we have addressed the problem of preparing a quantum many-body thermal state via a variational quantum algorithm suitable for NISQ computers. We utilized a parametrized quantum circuit composed of two blocks: the former acting on an ancillary register tasked with reproducing the Boltzmann distribution, with the latter acting on the system register tasked with rotating the system's state from the computational basis to the eigenenergy basis. The two registers are connected by a set of \textsc{CNOT} gates, and the optimizer is tasked with minimizing the generalized Helmholtz energy in order to prepare the Gibbs state on the system register.

We have improved on the Grover and Rudolph procedure for the preparation of an arbitrary probability distribution function, by reducing the number of exponential parameters needed in the VQA exploiting the symmetries of the $XY$ spin-$\frac{1}{2}$ Heisenberg model. We have then applied our protocol to a four-site chain and showed that our VQA is able to reach almost unit fidelity in statevector simulations by combining the Grover and Rudolph algorithm, with a reduced number of variational parameters acting on the ancillary register, and a parity-preserving ansatz on the system register. Although we have applied our protocol to small system sizes, our results hint toward the fact that a sub-exponential number of variational parameters are necessary for the preparation of a many-body Gibbs distribution by exploiting the degeneracies in the energy spectrum.

\section*{Acnowledgements}

The Authors acknowledge discussions with Francesco Plasina and Jacopo Settino on the topic of this chapter. TJGA ackowledges funding through the IPAS+ (Internationalisation Partnership Awards Scheme +) QIM project by the MCST (The Malta Council for Science \& Technology). M.C. acknowledges funding by TESS (Tertiary Education Scholarships Scheme).

\bibliographystyle{unsrt}
\bibliography{ref.bib}

\begin{thebibliography}{10}

\bibitem{Bengtsson2006}
Ingemar Bengtsson and Karol Zyczkowski.
\newblock {\em {Geometry of Quantum States: An Introduction to Quantum
  Entanglement}}.
\newblock Cambridge University Press, 2006.

\bibitem{Barahona1982}
F.~Barahona.
\newblock {On the computational complexity of Ising spin glass models}.
\newblock {\em Journal of Physics A: Mathematical and General},
  15(10):3241--3253, oct 1982.

\bibitem{Harrow2017}
Aram~W. Harrow and Ashley Montanaro.
\newblock {Quantum computational supremacy}.
\newblock {\em Nature}, 549(7671):203--209, sep 2017.

\bibitem{Preskill2018quantumcomputingin}
John Preskill.
\newblock Quantum {C}omputing in the {NISQ} era and beyond.
\newblock {\em {Quantum}}, 2:79, August 2018.

\bibitem{Bharti2022}
Kishor Bharti, Alba Cervera-Lierta, Thi~Ha Kyaw, Tobias Haug, Sumner
  Alperin-Lea, Abhinav Anand, Matthias Degroote, Hermanni Heimonen, Jakob~S.
  Kottmann, Tim Menke, Wai-Keong Mok, Sukin Sim, Leong-Chuan Kwek, and Al\'an
  Aspuru-Guzik.
\newblock {Noisy intermediate-scale quantum algorithms}.
\newblock {\em Rev. Mod. Phys.}, 94:015004, Feb 2022.

\bibitem{Cerezo2020}
M.~Cerezo, Andrew Arrasmith, Ryan Babbush, Simon~C. Benjamin, Suguru Endo,
  Keisuke Fujii, Jarrod~R. McClean, Kosuke Mitarai, Xiao Yuan, Lukasz Cincio,
  and Patrick~J. Coles.
\newblock {Variational quantum algorithms}.
\newblock {\em Nature Reviews Physics}, 3(9):625--644, aug 2021.

\bibitem{Tilly2022}
Jules Tilly, Hongxiang Chen, Shuxiang Cao, Dario Picozzi, Kanav Setia, Ying Li,
  Edward Grant, Leonard Wossnig, Ivan Rungger, George~H. Booth, and Jonathan
  Tennyson.
\newblock {The Variational Quantum Eigensolver: A review of methods and best
  practices}.
\newblock {\em Physics Reports}, 986:1--128, nov 2022.

\bibitem{Ratini2022}
Leonardo Ratini, Chiara Capecci, Francesco Benfenati, and Leonardo Guidoni.
\newblock {Wave Function Adapted Hamiltonians for Quantum Computing}.
\newblock {\em Journal of Chemical Theory and Computation}, 18(2):899--909, feb
  2022.

\bibitem{Consiglio2022}
Mirko Consiglio, Wayne~J Chetcuti, Carlos Bravo-Prieto, Sergi Ramos-Calderer,
  Anna Minguzzi, Jos{\'{e}}~I Latorre, Luigi Amico, and Tony J~G Apollaro.
\newblock {Variational quantum eigensolver for SU(N) fermions}.
\newblock {\em Journal of Physics A: Mathematical and Theoretical},
  55(26):265301, jul 2022.

\bibitem{Consiglio2022a}
Mirko Consiglio, Tony J~G Apollaro, and Marcin Wie{\'{s}}niak.
\newblock {Variational approach to the quantum separability problem}.
\newblock {\em Physical Review A}, 106(6):062413, dec 2022.

\bibitem{Watrous2008}
John Watrous.
\newblock {Quantum Computational Complexity}, 2008.

\bibitem{Holmes2022}
Zoe Holmes, Gopikrishnan Muraleedharan, Rolando~D. Somma, Yigit Subasi, and
  Burak {\c{S}}ahino{\u{g}}lu.
\newblock {Quantum algorithms from fluctuation theorems: Thermal-state
  preparation}.
\newblock {\em {Quantum}}, 6:825, October 2022.

\bibitem{Motta2020}
Mario Motta, Chong Sun, Adrian T.~K. Tan, Matthew~J. O'Rourke, Erika Ye,
  Austin~J. Minnich, Fernando G. S.~L. Brand{\~a}o, and Garnet Kin-Lic Chan.
\newblock {Determining eigenstates and thermal states on a quantum computer
  using quantum imaginary time evolution}.
\newblock {\em Nature Physics}, 16(2):205--210, Feb 2020.

\bibitem{Haug2022}
Tobias Haug and Kishor Bharti.
\newblock {Generalized quantum assisted simulator}.
\newblock {\em Quantum Science and Technology}, 7(4):045019, aug 2022.

\bibitem{Schaller2008}
Gernot Schaller.
\newblock {Adiabatic preparation without quantum phase transitions}.
\newblock {\em Physical Review A}, 78(3):032328, sep 2008.

\bibitem{Consiglio2023a}
Mirko Consiglio, Jacopo Settino, Andrea Giordano, Carlo Mastroianni, Francesco
  Plastina, Salvatore Lorenzo, Sabrina Maniscalco, John Goold, and Tony J.~G.
  Apollaro.
\newblock Variational gibbs state preparation on nisq devices, 2023.

\bibitem{Grover2002}
Lov Grover and Terry Rudolph.
\newblock Creating superpositions that correspond to efficiently integrable
  probability distributions, 2002.

\bibitem{LIEB1961407}
Elliott Lieb, Theodore Schultz, and Daniel Mattis.
\newblock {Two soluble models of an antiferromagnetic chain}.
\newblock {\em Annals of Physics}, 16(3):407--466, 1961.

\bibitem{Damski2014}
Bogdan Damski and Marek~M Rams.
\newblock {Exact results for fidelity susceptibility of the quantum Ising
  model: the interplay between parity, system size, and magnetic field}.
\newblock {\em Journal of Physics A: Mathematical and Theoretical},
  47(2):025303, jan 2014.

\bibitem{Franchini2016a}
Fabio Franchini.
\newblock {\em {An Introduction to Integrable Techniques for One-Dimensional
  Quantum Systems}}, volume 940 of {\em Lecture Notes in Physics}.
\newblock Springer International Publishing, Cham, sep 2017.

\bibitem{Herbert2021}
Steven Herbert.
\newblock No quantum speedup with grover-rudolph state preparation for quantum
  monte carlo integration.
\newblock {\em Phys. Rev. E}, 103:063302, Jun 2021.

\bibitem{Nakaji2022}
Kouhei Nakaji, Shumpei Uno, Yohichi Suzuki, Rudy Raymond, Tamiya Onodera,
  Tomoki Tanaka, Hiroyuki Tezuka, Naoki Mitsuda, and Naoki Yamamoto.
\newblock Approximate amplitude encoding in shallow parameterized quantum
  circuits and its application to financial market indicators.
\newblock {\em Phys. Rev. Res.}, 4:023136, May 2022.

\bibitem{Dasgupta2022}
Kalyan Dasgupta and Binoy Paine.
\newblock Loading probability distributions in a quantum circuit, 2022.

\bibitem{Kumar2023}
Rohit~Taeja Kumar and Ankur Raina.
\newblock Generating probability distributions using variational quantum
  circuits, 2023.

\bibitem{Zoufal2019}
Christa Zoufal, Aur{\'e}lien Lucchi, and Stefan Woerner.
\newblock Quantum generative adversarial networks for learning and loading
  random distributions.
\newblock {\em npj Quantum Information}, 5(1):103, Nov 2019.

\bibitem{Situ2020}
Haozhen Situ, Zhimin He, Yuyi Wang, Lvzhou Li, and Shenggen Zheng.
\newblock Quantum generative adversarial network for generating discrete
  distribution.
\newblock {\em Information Sciences}, 538:193--208, 2020.

\bibitem{Agliardi2022}
Gabriele Agliardi and Enrico Prati.
\newblock Optimal tuning of quantum generative adversarial networks for
  multivariate distribution loading.
\newblock {\em Quantum Reports}, 4(1):75--105, 2022.

\bibitem{Benenti2009}
Giuliano Benenti and Giuliano Strini.
\newblock Optimal purification of a generic $n$-qudit state.
\newblock {\em Phys. Rev. A}, 79:052301, May 2009.

\bibitem{Consiglio2023b}
Mirko Consiglio.
\newblock Variational quantum algorithms for gibbs state preparation, 2023.

\bibitem{Uhlmann2011}
Armin Uhlmann.
\newblock {Transition Probability (Fidelity) and Its Relatives}.
\newblock {\em Foundations of Physics}, 41(3):288--298, 2011.

\end{thebibliography}

\end{document}